\documentclass[12pt]{iopart}

\eqnobysec

\usepackage{latexsym,amssymb}	
\usepackage[dvips]{epsfig}	

\newcommand{\ind}[1]{{\mbox{\scriptsize #1}}}

\begin{document}
\jl{6}

\title[Gravitational Scattering of  Cosmic Strings]
{Gravitational Scattering of Cosmic Strings by 
Non-Rotating Black Holes.}

\author{Jean-Pierre De Villiers\dag
\footnote{e-mail: jpd@phys.ualberta.ca}
 and Valeri Frolov\ddag
\footnote{e-mail: frolov@phys.ualberta.ca}}

\address{\dag\ Department of Astronomy and Physics, Saint Mary's 
University, Halifax, Nova Scotia, Canada B3H 3C3}

\address{\ddag\ Theoretical Physics Institute, Department of Physics, 
University of Alberta, Edmonton, Canada T6G 2J1}

\begin{abstract}  
\noindent 
This paper discusses the gravitational scattering of a  straight,
infinitely long test cosmic string   by a black hole. We present
numerical results that probe the two-dimensional parameter space of
impact parameter and initial velocity and compare them to approximate
perturbative solutions derived previously.  We analyze string
scattering and coil formation in the ultra-relativistic regime and
compare these results with analytical results for string
scattering by a gravitational shock wave.  Special attention is paid to
regimes  where the string approaches the black  hole at
near-critical impact parameters.  The dynamics of string scattering in this
case are  highly sensitive to initial data and transient
phenomena arise while portions of the string dwell in the strong
gravitational field near the event horizon of the black  hole. 
The role of string tension is also
investigated by comparing the scattering of a cosmic string to
the scattering of a tensionless "dust" string. Finally, the
problem of string capture is revisited in light of these new
results, and a capture curve  covering  the entire velocity range ($0 <
v \le c$) is given. 
\end{abstract}

{\it PACS number(s): 04.60.+n, 12.25.+e, 97.60.Lf, 11.10.Gh}

\submitted

\maketitle

\newpage
\baselineskip=.8cm

\section{Introduction}
\setcounter{equation}0

In this paper we study the interaction of a cosmic string with a
(non-rotating) black hole. This problem is interesting for many
reasons. First of all, it is a rare example of  gravitational 
interaction between two extended relativistic objects that allows
complete analysis. When a cosmic string comes close to a black hole and
is captured by it, one might expect generation of strong gravitational
radiation, which makes the problem potentially interesting for
astrophysical applications. Besides this, the study of the interaction
of a cosmic string with a black hole has demonstrated some unexpected,
physically interesting possibilities, such as the creation of 2D black
holes \cite{FrHeLa:96}. 

For astrophysically interesting cases, the dimensionless  parameter
$\mu^*=G\mu/c^2$, where $\mu$ is the string tension, is small. For
example, for strings formed during a GUT phase transition,
$\mu^*\approx 10^{-6}$. For this reason, in the leading order one can
neglect gravitational backreaction  effects and consider the string as
a test object propagating in a given gravitational background. The
worldsheet of such a string is a minimal surface which gives an
extremum to the Nambu-Goto action \cite{Nambu:69,Goto:71}.

Stationary configurations of a cosmic string in the field of a rotating
charged black hole are described by worldsheets which are tangent to
the Killing vector generating the time shift. Because of this 
symmetry, the problem is reduced to finding a geodesic in a
three-dimensional space;  the latter  can be solved by separation of
variables \cite{FrSkZeHe:89}.  In the absence of such a symmetry,
non-linear partial differential equations describing the string motion
in the gravitational field of the black hole do not allow exact
solutions. Under these conditions one either needs approximation
schemes or numerical solutions.
In the present paper we assume that the string is initially far from
the non-rotating black hole, straight, and moving with velocity $v$ in the 
direction of the black hole. The  motion of the string is affected
by the gravitational field of the black hole.  For a given velocity
$v$, the outcome of the interaction  depends on the impact parameter
${b}$. If the impact parameter is large enough,  the string escapes the
gravitational attraction of the black hole and is inelastically
scattered. For smaller values of ${b}$ the string is captured by the
black hole. 

The scattering of a cosmic string for very large impact parameters was
studied in \cite{DVFr:98,Page:98}. Such a string passes far from the
black hole and its motion is modified only slightly. Under these
conditions one can consider the gravitational interaction as a
perturbation and treat it by expanding the solution in powers of the
Newtonian gravitational potential $\varphi=GM/(Rc^2)$. The numerical study
of string scattering by a black hole was initiated by Moss and Lonsdale
\cite{LoMo:88}. In Reference \cite{DVFr:97} we used numerical calculations
to obtain the dependence of the critical capture impact parameter on
velocity $v$, and demonstrated  that the results of Moss and Lonsdale
do not correctly reproduce the behaviour of the critical impact
parameter at high velocities.

In order to obtain more detailed information on the characteristics of
scattering and capture of a cosmic string by a black hole we
performed further numerical calculations.  In regimes that allow
analytical study, we compared  numerical and analytical  results. The
agreement of these results was used as an additional test of our
calculations. We also analysed the role of tension in the string by
comparing the relativistic  string results with those for a string
without pressure (a so-called ``dust model'').  These results are
discussed in the following sections.  In Section~2 we  formulate the 
scattering problem for
the straight string. Section~3 outlines the numerical method that was
used and discusses the scattering at shallow impact parameters. It also
contains comparison of these results with  the results obtained in the
weak-field approximation. Section~4 compares numerical and approximate
analytical results for a string moving with ultra-relativistic
velocity.  It also discusses coil formation for the scattering in this
regime. (We make the distinction here 
between coils, which are loop-like configurations in a straight string, 
and the term loop, as found in the literature, which designates closed, bounded
string configurations. We thank an anonymous referee for suggesting this
clarification.) Section~5 discusses near-critical scattering, where the string
worldsheet is partially wrapped around the black hole. Section~6
discusses the role played by the tension of the string and makes a
comparison with the model of a dust string. Section~7  discusses
aspects of the problem connected with string capture. 

\section{Scattering Problem for a  Straight String}
\setcounter{equation}0

\subsection{Equations of motion}

The two-dimensional worldsheet representing  a moving string is
parametrized as ${\cal X}^{\mu}(\zeta^A)$; ${\cal X}^{\mu}$
($\mu=0,1,2,3$) are the spacetime coordinates, and $\zeta^A$ ($A=0,3$)
are the worldsheet coordinates $\zeta^0=\tau$, $\zeta^3=\sigma$.  The 
equations of motion of the string follow from the Nambu-Goto action
\cite{ShVi:94} which we write in the Polyakov form \cite{Poly:81}
\begin{equation}\label{2.1} 
I[{\cal X},h]=-{\mu \over 2}\,\int d\tau d\sigma \sqrt{-h}h^{AB}G_{AB}\,
.
\end{equation}
We use units in which $G=c=1$, and the
sign conventions of \cite{MTW}.  
In (\ref{2.1})  $h_{AB}$ is the internal metric
with determinant $h$, and $G_{AB}$ is the induced metric on the
worldsheet,
\begin{equation}\label{2.2} 
G_{AB}=g_{\mu\nu}{\partial {\cal X}^{\mu}\over \partial\zeta^A}{\partial
{\cal X}^{\nu}\over \partial\zeta^B}=g_{\mu\nu} {\cal X}^{\mu}_{,A}{\cal
X}^{\nu}_{,B} \, ,
\end{equation}
where $g_{\mu\nu}$ is the spacetime metric.

The equations obtained by varying  action (\ref{2.1}) with respect to
${\cal X}^{\mu}$ and $h_{AB}$ are of the form
\begin{equation}\label{2.3} 
\Box {\cal X}^{\mu}+h^{AB}\Gamma^{\mu}_{\alpha\beta}{\cal
X}^{\alpha}_{,A}{\cal X}^{\beta}_{,B}=0
\end{equation}
and
\begin{equation}\label{2.4} 
G_{AB}-{1\over 2}h_{AB}h^{CD}G_{CD}=0 \, ,
\end{equation}
where
$\Box =(-h)^{-1/2}\partial_A\left[(-h)^{-1/2}h^{AB}\partial_B\right]$.
Equations (\ref{2.3}) are the dynamical equations for string
motion, while equations (\ref{2.4}) are constraints. 

\subsection{Straight string motion in flat spacetime}

Consider first the motion of a straight cosmic string in the absence
of an external gravitational field, that is when
$g_{\mu\nu}=\eta_{\mu\nu}$, where $\eta_{\mu\nu}$ is the flat spacetime
metric. In Cartesian coordinates ($T,X,Y,Z$),
$\eta_{\mu\nu}=\mbox{diag}(-1,1,1,1)$ and
$\Gamma^{\mu}_{\alpha\beta}=0$, and it is easy to verify that,
taking $h_{AB}=\eta_{AB}\equiv \mbox{diag}(-1,1)$,
\begin{equation}\label{2.5} 
{\cal X}^{\mu}=X^{\mu}(\tau,\sigma)\equiv (\cosh(\beta)\, \tau,
\sinh(\beta)\,\tau+X_0, Y_0,\sigma) 
\end{equation}
is a solution of equations (\ref{2.3}) and (\ref{2.4}). This solution
describes a straight string oriented along the $Z$-axis which moves in
the $X$-direction with constant velocity $v=\tanh \beta$. Initially, at
${\tau}_{0} = 0$, the string is found at  ${\cal X}^{\mu}(0,\sigma) =
(0,X_0, Y_0,\sigma)$, where $Y_0 \equiv b$ is the impact parameter. 
For definiteness we choose $Y_0>0$ and $X_0<0$,
so that $\beta>0$.

The two-dimensional surface of the
worldsheet of such a string is spanned by vectors 
\begin{equation}\label{2.7} 
e_{(0)}^{\mu}=X_{,\tau}=(\cosh\beta, \sinh\beta, 0, 0)
\end{equation}
and
\begin{equation}\label{2.7a} 
e_{(3)}^{\mu}=X_{,\sigma}=(0, 0, 0, 1) \, ,
\end{equation}
while the vectors 
\begin{equation}\label{2.8}
e_{(1)}^{\mu}=n_1^{\mu}=(\sinh\beta, \cosh\beta, 0, 0) 
\end{equation}
and
\begin{equation}\label{2.8a} 
e_{(2)}^{\mu}=n_2^{\mu}=(0, 0, 1, 0) 
\end{equation}
are orthogonal to the worldsheet.
Vectors  $e^{\mu}_{(m)}$ ($m=0,1,2,3$) form an  orthogonal tetrad.

\subsection{String motion in a weak gravitational field}

The study of scattering of an infinitely long cosmic string by a
non-rotating  black hole is most conveniently carried out in isotropic
coordinates, $(T,X,Y,Z)$, for which the Schwarzschild metric has the
form
\begin{equation}\label{2.9} 
d{s}^{2}= 
-{{\left(1 - M/2R \right)}^{2}\over {\left(1 + M/2R
\right)}^{2}}\,d{T}^{2} +
{\left(1 + {M \over 2\,R}
\right)}^{4}\,\left(d{X}^{2}+d{Y}^{2}+d{Z}^{2}\right)\, ,
\end{equation} 
where $R^2=X^2+Y^2+Z^2$. This metric can be written as
\begin{equation}\label{2.10} 
ds^2=-(1-2\Phi)dT^2+(1+2\Psi)(dX^2+dY^2+dZ^2)\, 
\end{equation}
with
\begin{equation}\label{2.11} 
\Phi={\varphi\over (1+{1\over 2}\varphi)^2}
\end{equation}
and
\begin{equation}\label{2.11a} 
\Psi=\varphi+{3\over 4}\varphi^2+{1\over 4}\varphi^3 +
{1\over 32}\varphi^4\,,
\end{equation}
where $\varphi = M/R$ is the Newtonian potential. 

\begin{figure}[ht]
\centerline{\epsfxsize=8cm \epsfbox{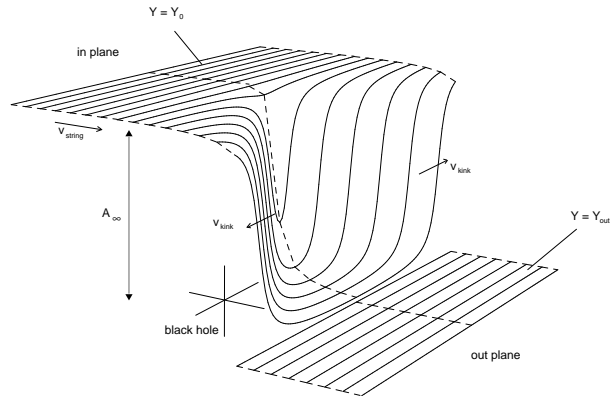}}
\caption{A straight cosmic string scattered by a Schwarzschild black hole.}
\end{figure}

The motion of the string in the region with small $\varphi$ can be
written in the form
\begin{equation}\label{2.12} 
{\cal X}^{\mu}(\tau,\sigma)=X^{\mu}(\tau,\sigma)+\delta
X^{\mu}(\tau,\sigma)\, .
\end{equation}
The straight string starts its motion in the plane $Y=Y_0 = b$. We call it
``in-plane'' (see Figure~1). 
This plane represents the motion of the free string in flat spacetime. 
Deflection of the string worldsheet from the ``in-plane'', $\delta X^{\mu}$,
can be decomposed in the orthogonal frame as
\begin{equation}\label{2.13} 
\delta X^{\mu}=\chi^{(m)} e^{\mu}_{(m)} \, .
\end{equation}
By linearizing  string equations (\ref{2.3}) and (\ref{2.4}) one gets
the following equations for the first-order perturbations $\chi^{(m)}$
\cite{DVFr:98}:
\begin{equation}\label{2.14} 
\Box {\chi}_{(m)}=\left(-{\partial^2\over \partial\tau^2}+ {\partial^2\over
\partial\sigma^2}\right){\chi}_m={f}_m\, ,
\end{equation}
where
\begin{equation}\label{2.16} 
{f}_{(0)}=2\sinh\beta\, \cosh^2\beta \, \, \varphi_{,X}\, ,
\end{equation}
\begin{equation}\label{2.15} 
{f}_{(1)}=2\sinh^2\beta\, \cosh\beta \, \, \varphi_{,X}\, ,
\end{equation}
\begin{equation}\label{2.15a} 
{f}_{(2)}=-2\sinh^2\beta\,\, \varphi_{,Y}\, ,
\end{equation}
and
\begin{equation}\label{2.16a} 
{f}_{(3)}=-2\cosh^2\beta\,\, \varphi_{,Z}
\, .
\end{equation}

The solutions of  linearized equations (\ref{2.14})--(\ref{2.16a})
were obtained and analyzed in 
\cite{DVFr:98}. Here we reproduce only a solution for ${\chi}_{(2)}$ which
describes the deflection of the string in $Y$-direction, that is in the
spatial direction perpendicular to the string motion. We assume that a straight
string starts its motion with $\chi_{(2)}=0$ at $X_0=-\infty$ then
\begin{equation}\label{2.17}
{\chi}_2(\tau,\sigma)=-M\sinh\beta\,\left[H_{+}(\tau,\sigma)+
H_{-}(\tau,\sigma)\right]
\, ,
\end{equation}
where
\begin{equation}\label{2.18}
H_{\pm}(\tau,\sigma)=
\arctan{\left[{{Y}_{0}^{2} + \tau (\tau\pm\sigma)\,\sinh^2\beta}
\over{Y_0 \,R\,\sinh\beta}\right]}-
\arctan{\left[{(\tau\pm\sigma)\,\tanh\beta}\over{Y_0}\right]}
\, .
\end{equation}
Here $R^2=\tau^2\sinh^2\beta +Y_0^2+\sigma^2$. 
%
This expression for ${\chi}_2(\tau,\sigma)$ can be easily obtained from
equation (2.22) of Ref.\cite{DVFr:98} if we change $\tau \rightarrow
\tau - X_0/\sinh \beta$ in this expression and take the limit $X_0
\rightarrow -\infty$. After this redefinition of $\tau$,  the value of
new variable $\tau=0$ corresponds to the moment of time when the
``distance'' $R$ from the string to the black hole is minimal.
%
%

The analysis of the late time asymptotic of this solution shows that 
 the string approaches the
plane  $Y=Y_{out}$, and the shift in position of the string is given by
\footnote{When $v\rightarrow 0$ second order corrections become
important (see \cite{DVFr:98,Page:98}). We do not consider the case of
extremely slow motion of strings in this paper and always assume that
$v>0.01c$. 
For a discussion of slowly moving strings see \cite{Page:99}.
}
\begin{equation}\label{2.19}
{A}_{\infty} = Y_{0}- Y_{out}=  2 M\,\pi\,\sinh{\beta}=
2 M\,\pi\,\gamma\,v,
\end{equation}
%
%
where $v = \tanh{\beta}$ is the velocity of the string
and $\gamma=(1-v^2)^{-1/2}$. The late-time
%
%
solution is represented by a kink and anti-kink, propagating in
opposite directions at the speed of light, and leaving behind them the
string in a new ``phase''. Furthermore, it can be shown that each kink 
has a characteristic width in the external spacetime given 
by \footnote{Note that there is an additional factor of $\pi$ in the expression 
for the width of the kinks which was absent  in \cite{DVFr:98} .}
\begin{equation}\label{2.20}
w = \pi\,Y_0\,\coth{\beta} = \pi\,{Y_0 \over v}.
\end{equation}

\subsection{Scattering data}

The above results were obtained under the assumption that the string
always propagates in the region where the gravitational field 
remains weak.  In what follows, we describe results concerning 
string scattering in the regime where the gravitational field cannot be
considered weak.  Nevertheless, if the string is not captured it 
eventually emerges from the
strong field region and its subsequent evolution is described again by equations
(\ref{2.14}). The string's late time features are
characterized by the amplitude 
${A}_{\infty}$ and width $w$ of the kinks; we refer to these two
parameters as the {\em scattering data}.  As will be seen in the following
sections, the  breakdown of the weak field
approximation manifests itself in two ways: the failure  of the
analytical expressions to correctly reproduce transient shapes, which
develop when the string reaches periastron, and discrepancies in the
late-time features of the string, namely the scattering data.

\section{Strong Field Scattering}
\setcounter{equation}0

\subsection{Numerical methods} 

In order to study the scattering problem under strong-field conditions 
(where the weak-field approximation breaks down), a numerical solution 
to the equations of motion must be
found.  To deal with an infinite cosmic string numerically, a finite
computational domain is required, so a method of
truncating the  string to a reasonably short segment is needed. Such
a physical truncation  imposes special boundary conditions since it is
crucial to reproduce the motion of an infinitely long string with
sufficient accuracy.

Two methods of truncating a string were developed.  The first  places
a massive particle at each end of the segment of
string and requires that the motion of these particles mimic the motion
of the portions of the string lying outside the region of interest. This is
accomplished by letting the mass of the end particles go to infinity.
%
%
It should be emphasized  that truncating the string and adding infinitely
heavy particles at its ends is just a technical trick. It allows one to
consider finite strings. In making the size of the string longer and
longer, we finally reach a size where a
further increase in the string's length does not affect the solution
within the chosen accuracy. This
may be interpreted as saying that the infinitely massive end particles
correctly simulate an infinite string provided the length of the
truncated string is chosen sufficiently long
(this issue is discussed in Reference \cite{DVFr:97}).
%
%

The second method places the truncation points a reasonable distance
away from the region of interest, subject to the condition that the end
points remain in a weak gravitational field where a perturbative
solution to the equations of motion is applicable. The analytic
weak-field solution is used to prescribe the motion of the boundary
points of the string. 
Each method has distinct advantages, and both were used extensively. 
Further, the solvers were tested against one another and agreement 
under identical initial conditions served as an
additional validation of the numerical schemes.

With the choice of internal metric $h_{AB} ={e}^{2\Omega} {\eta}_{AB}$,
the  equations of motion have the form
\begin{equation}\label{3.1}
{{\partial}^{2}\,{\cal{X}}^{\mu} \over \partial\,{\tau}^{2}}-
{{\partial}^{2}\,{\cal{X}}^{\mu} \over \partial\,{\sigma}^{2}}+
{\Gamma}^{\mu}_{\rho \eta}\,\left(
{\partial\,{\cal{X}}^{\rho} \over \partial\,\tau}\,
{\partial\,{\cal{X}}^{\eta} \over \partial\,\tau}-
{\partial\,{\cal{X}}^{\rho} \over \partial\,\sigma}\,
{\partial\,{\cal{X}}^{\eta} \over \partial\,\sigma}\right) = 0 \, .
\end{equation}
The  discretization of these equations 
(described in detail in Reference  \cite{DVFr:97}) 
yields a block-tridiagonal system with which the
solution over the entire spatial  ($\sigma$) grid is obtained by
algebraic methods. The boundary conditions, for which 
${\cal{X}}^{\mu}(\tau,{\sigma}_{i}) \equiv {{X}}^{\mu}_{i}$, are
imposed by replacing the first and last entries in the block-tridiagonal 
system by  the required expressions. 

For infinitely massive end particles at the boundaries 
the following equations are substituted into the system,
\begin{equation}\label{3.2}
{{d}^{2}\,X^{\mu}_i \over d\,{\tau}^{2}}+
{\Gamma}^{\mu}_{\rho \eta}\,
{d\,{X}^{\rho}_i \over d\,\tau}\,
{d\,{X}^{\eta}_i \over d\,\tau} = 0\, ;
\end{equation}
the derivation of this boundary condition is
given in  Reference \cite{DVFr:97}.
For the perturbative boundary conditions, the weak-field solutions 
of Eqn.(\ref{2.14}) are substituted into the system; the solutions are given 
in Reference \cite{DVFr:98}.

The technical details of the numerical work are not important, but
a few general comments are in order on the practical aspects imposed
by the boundary conditions and initial data. 

Under both types of boundary conditions, 
the accuracy of the numerical results may be influenced by the length 
$L$ of the string segment, meaning that if a string segment is too short,
the motion of the boundary points induces incorrect behaviour. The
larger the value of $L$, the smaller the effect, but the larger the value of
$L$, the greater the number of grid points required to adequately resolve
the string. The choice of string length is therefore a compromise between
the need to accurately reproduce the behaviour of an infinite string and the
need to use computing resources effectively.  We found that the 
massive-particle boundary conditions required a string length of at least 
$L = 1000 {r}_{g}$ 
%
%
(${r}_{g} = 2 M$, the Schwarzschild radius of the black hole)
%
%
to give reasonable results, as discussed in Reference  
\cite{DVFr:97}, whereas the perturbative boundary conditions required a much 
shorter string length,  $L \sim 100\,{r}_{g}$, since the weak-field solutions 
represent the motion of the string boundary points with sufficient accuracy at 
these distances.  Therefore, the perturbative boundary conditions greatly 
reduce the  number of grid points and time steps needed to resolve the motion 
of the string  near the black hole, and hence speed up execution dramatically. 

Initial data takes the form of a straight string, Minkowski solution
with the stipulation that the initial position of the string is
sufficiently far away from the black hole to minimize the discrepancy
between this approximate  initial data and the correct initial
configuration in Schwarzschild spacetime; the constraint equations are
used as a guide.  The constraint equations are also used to check the
accuracy of the calculations during the time evolution of the string. 

The evaluation of the discretized
form of the constraint equations is carried out periodically
during the numerical solution. Statistics are computed for the constraints
(average value over the length of the string and standard deviation) at the 
current time step and reported to an
output file. Monitoring that the constraints are consistent with zero to several
significant digits is done by inspecting this file. If the constraints are
not satisfactorily maintained, that is if the average value grows unacceptably
large or undergoes sudden changes, the numerical solution is restarted with
new parameters (e.g. finer grid, finer step size). 
%
%
The results for the constraint 
calculation are influenced
significantly by the initial position of the string. A solution for a
string placed initially closer to the black hole requires fewer time steps,
but generates a poorer result in the constraint calculation.
Typically, the constraints were expected to be consistent with zero from
four to six significant figures. 
%
%
Tighter tolerances can also be 
achieved by increasing the number of grid points (and significantly increasing 
solution time), but for the purposes of this study, the above
values were deemed a reasonable compromise between accuracy and speed.

The perturbative solution to the equations of motion of an infinitely
long  cosmic string is most conveniently carried out in isotropic
coordinates,  $(T,X,Y,Z)$. Numerically, however, the scattering problem
is best studied in  Eddington-Finkelstein In-going coordinates since
the Christoffel symbols in this  coordinate system are  regular
everywhere away from the origin. The Christoffel symbols associated with this coordinate 
system,  along with  their derivatives with respect to the spacetime
coordinates,  were derived and inserted as analytic expressions in the
portions of the code dealing with the initialization of the tridiagonal
matrices. Also, the coded forms of the analytic expressions 
for the initial data and for the perturbative boundary conditions were 
converted to Eddington-Finkelstein
coordinates using standard coordinate transformations. 

\subsection{Scattering at intermediate velocities}
\setcounter{equation}0

In this section, we compare the numerical solutions of string
scattering to those of the weak-field approximation at intermediate
velocities ($0.1 < v/c < 1$) for a range of impact parameters.
Our basis for comparison is the scattering data.

Figure~2 shows sections of the numerical worldsheet of a string with
initial velocity $0.76c$ and impact parameter $40\,{r}_{g}$. As shown in 
Table~1,
the maximum amplitude and width of 
the pulses is in good agreement with the weak-field 
Eqns.~(\ref{2.19}) and (\ref{2.20}).
\begin{table}[ht]
\begin{center}
\caption{Scattering data - Schwarzschild weak-field.}
\begin{tabular}{lll}
\hline
 & & \\
 & numerical & perturbation \\
\hline
 & & \\
 $A_{\infty}$  & $3.4\,{r}_{g}$ & $3.6\,{r}_{g}$\\
 $w$   & $140 {r}_{g}$  & $165 {r}_{g}$  \\
 & &  \\
\hline
& ($v = 0.76\,c$, & $b = 40\,r_g$)
\end{tabular}
\end{center}
\end{table}

\begin{figure}[ht]
\centerline{\epsfxsize=9cm \epsfbox{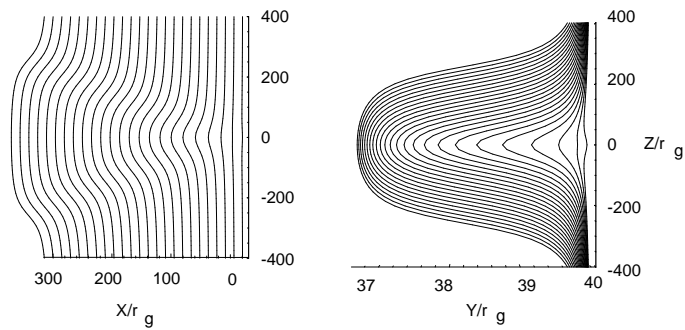}}
\caption{Time sequence of string scattering in weak-field regime 
(numerical results). Black hole lies at origin of coordinate system. 
Initial velocity 0.76c, impact parameter 40 ${r}_{g}$.}
\end{figure}

When the impact parameter approaches the critical value for 
capture, the string is in a regime where only numerical solutions are 
available. Figure~3 shows sections of the worldsheet of a slow string 
($v = 0.2 c$) with near-critical impact parameter ($b = 2.5$ ${r}_{g}$ 
compared to ${b}_{\ind{capture}} = 2.1$  ${r}_{g}$). As shown in Table~2,
the maximum amplitude of the pulses is considerably greater 
than predicted by weak-field Eqn.~(\ref{2.19}), while there is
good agreement in terms of the measured width of the kink and 
Eqn.~(\ref{2.20}).
\begin{table}[ht]
\begin{center}
\caption{Scattering data - Schwarzschild strong-field.}
\begin{tabular}{lll}
\hline
 & & \\
 & numerical & perturbation \\
\hline
 & & \\
 $A_{\infty}$ &  $1.8 \,r_g$ & $0.95\,r_g$ \\
 $w$  &  $28 \, r_g$      & $27\,r_g$   \\
 & &  \\
\hline
& ($v = 0.29\,c$,& $b = 2.5\,r_g$)
\end{tabular}
\end{center}
\end{table}

\begin{figure}[ht]
\centerline{\epsfxsize=10cm 
\epsfbox{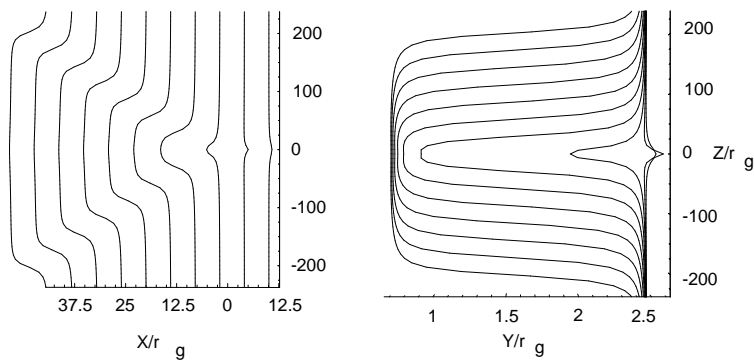}}
\caption{Time sequence of string scattering in strong-field regime 
(numerical results). Black hole lies at origin of coordinate system. 
Initial velocity, 0.29c, impact parameter 2.5 ${r}_{g}$.}
\end{figure}

It is important to note that, although the behaviour of the string is
qualitatively consistent with the weak-field solutions, the total
deflection  of the string is greater than predicted by
Eqn.~(\ref{2.19}). The discrepancy is made clear by plotting the total
deflection of the string obtained from the numerical solver and
comparing it to the prediction of Eqn.~(\ref{2.19}). This  is done in
Figure ~4, where the ratio ${A}_{numerical}/{A}_{weak}$ is plotted  for
four velocities. It can be seen that numerical and perturbative
results  converge for large impact parameters, and that the transition
from weak-field  to strong-field occurs for impact parameters on the
order of $10\,{r}_{g}$.  Conversely, the curves make it clear that the
weak-field solutions are quite  acceptable down to very small impact
parameters. The curve which suffers a  downturn is the lower velocity
one, meaning that Eqn.~(\ref{2.19})  predicts a greater shift than the
numerical data. The velocity of the string  in this case is at the
upper threshold where second-order corrections to the  perturbative
expansion become important. Recall that the expression  (\ref{2.19})
was  derived using results from first-order perturbative calculations.

\begin{figure}
\centerline{\epsfxsize=10cm 
\epsfbox{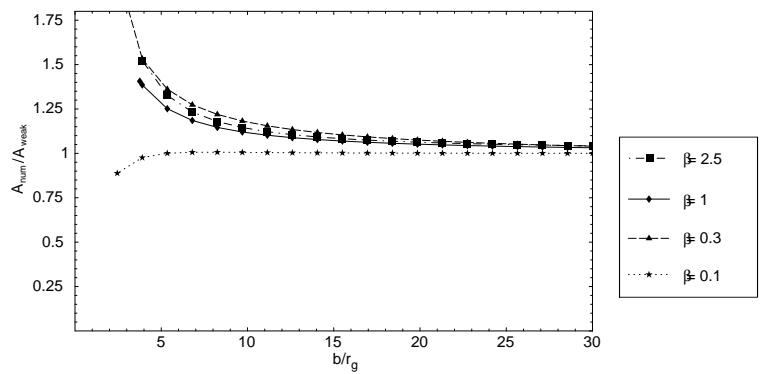}}
\caption{Comparison of numerical scattering date with weak-field
approximation.}
\end{figure}

\section{Ultra-relativistic Limit and Coil Formation}
\setcounter{equation}0

\subsection{Analytical results for ultra-relativistic scattering}

In this section we consider scattering of a straight  cosmic string by
a black hole in the limit where the initial velocity is close to the
velocity of light.  Numerical studies of the evolution of stochastic
networks of cosmic strings from initial formation to the matter era
\cite{ShVi:94} suggest that astrophysical cosmic strings are expected
to move with an average  velocity $v \approx 0.7 c$ . For this reason,
the results  discussed in this section concerning motion at the
ultra-relativistic limit are mainly of theoretical interest.
Nevertheless, they provide a number of  analytic expressions against
which to validate numerical solutions in this regime.

In the limit $v\rightarrow c$, the gravitational field of a black hole
in the reference frame of the string takes the form of a shock wave.
The corresponding  Aichelburg-Sexl metric \cite{AiSe:71} is of the form
(see Appendix)
\begin{equation}\label{4.1} 
ds^2=-dX_-\, dX_+ +dY^2+dZ^2 -4\gamma M F \delta(X_+) dX_+^2
\, ,
\end{equation}
where $F=\ln \rho^2$, and $\gamma=(1-v^2)^{-1/2}$. In these null
coordinates, the initial motion of the straight string is given by
\begin{equation}\label{4.2} 
\left. X_+\right|_{{\tau}<0}={\tau}\, ,
\end{equation}
\begin{equation}\label{4.2a} 
\left. X_-\right|_{{\tau}<0}={\tau}-2 \gamma X_0\, ,
\end{equation}
\begin{equation}\label{4.2b} 
\left. Y\right|_{{\tau}<0}=Y_0\, ,
\end{equation}
and
\begin{equation}\label{4.2c} 
\left. Z\right|_{{\tau}<0}=\sigma\,.
\end{equation}
The problem of scattering of the string in the metric (\ref{4.1}) was
studied earlier (see e.g. \cite{AmKl:88}). For our
case, the corresponding solutions are  (details in the Appendix)
\begin{equation}\label{4.3} 
\left. X_+\right|_{\tau>0}={\tau} \, ,
\end{equation}
%
\[ 
\left. X_-\right|_{{\tau}>0}=
\tau -2\gamma X_0 -
2\gamma M \left[\ln \left( 1+\left(\tau+\sigma\over
Y_0\right)^2\right)+\ln \left( 1+\left(\tau-\sigma\over
Y_0\right)^2\right)\right]
\]
\begin{equation}\label{4.5}
\quad\quad\quad +
{8\gamma^2 M^2\over Y_0}\left[ \arctan\left( \tau+\sigma\over Y_0\right)+
\arctan\left( \tau-\sigma\over Y_0\right)\right]
\, ,
\end{equation}
\begin{equation}\label{4.4} 
\left. Y\right|_{{\tau}>0}=Y_0-2\gamma M\, 
\left[\arctan\left({{\tau}+\sigma\over Y_0}\right)
+\arctan\left({{\tau}-\sigma\over Y_0}\right)
\right] \, ,
\end{equation}
and
\begin{equation}\label{4.3a} 
\left. Z\right|_{{\tau}>0}=\sigma-\gamma M\,\ln 
\left[{Y_0^2+({\tau}+\sigma)^2\over Y_0^2+({\tau}-\sigma)^2}
\right] \, .
\end{equation}

These results allow one to easily determine 
the amplitude of the kinks for the ultra-relativistic solutions.
The maximum amplitude comes from the asymptotic value of 
$\left. Y\right|_{\tilde{\tau}>0}$,
%
%
\begin{equation}\label{4.6}
A_{\infty} :=Y_0-
\lim_{\tilde{\tau} \rightarrow \infty} \left. Y\right|_{\tilde{\tau}>0} 
= 2 \pi\,\gamma\,M \, .
\end{equation}
%
%
This result is in agreement with the result of weak field approximation
(\ref{2.19}), in the limit $v \rightarrow c$.

\subsection{Coil formation}

The formation of coils characterises 
ultra-relativistic scattering. 
A coil-like configuration appears when $Z(\tau,\sigma)$ is no longer 
a monotonic function of $\sigma$ for a fixed value of $\tau$. This
%
%
occurs when $Z_{,\sigma}<0$. From Eqn.~(\ref{4.3a}), one gets
%
%
\begin{equation}\label{4.7} 
Z_{,\sigma}=1-2\gamma M\left[{\zeta_+\over Y_0^2+\zeta_+^2}+{\zeta_-\over
Y_0^2+\zeta_-^2}  \right]\, ,
\end{equation}
where $\zeta_{\pm}={\tau}\pm\sigma$. The function $Z_{,\sigma}$
has extrema at $|\zeta_{\pm}|=|Y_0|$. The minimum of $Z_{,\sigma}$
occurs at $\zeta_{+}=\zeta_{-}=Y_0$, that is at $\tau=Y_0$ and
$\sigma=0$. This minimal value  $Z_{,\sigma}=1-2\gamma M/Y_0$ becomes
negative when 
\begin{equation}\label{4.8}
2\gamma M>Y_0\, . 
\end{equation}
This is the condition for coil formation. This condition can be overlaid
on the capture curve \cite{DVFr:97}, as shown in Figure ~5. 
The coil formation region
lies above the capture curve and to the right of the curve 
$\gamma = {Y}_{0}/{r}_{g}$; it can be seen that coil formation
is a relativistic phenomenon, and the ultra-relativistic solution predicts 
that this effect cuts off for velocities below $v \sim 0.9 c$

The maximum width of the coils, $w_{coil}$, 
occurs when the solution 
for $\left. Z\right|_{\tilde{\tau}>0}$ reaches an extremum in both $\tau$ 
and $\sigma$. 
%
%
The maximum width represents the greatest distance,
relative to the $Z = 0$  plane, reached by the segment of string
forming the coil before the coil begins to "collapse" and
unwind (see Figure 6).
%
%
It is straightforward to compute the derivatives from 
Eqn.~(\ref{4.3a}) and show that
\begin{equation}\label{4.9}
w_{coil} = 2 \gamma\,\left\{\eta - {1 \over 2}
\ln \left[{1 + \eta \over 1 - \eta}\right]
\right\} (r_g)\quad ; 
\quad \eta = \sqrt{1 - \left({Y_0 \over 2 M \, \gamma}\right)^2}\,.
\end{equation}


\begin{figure}[ht]
\centerline{\epsfxsize=12cm 
\epsfbox{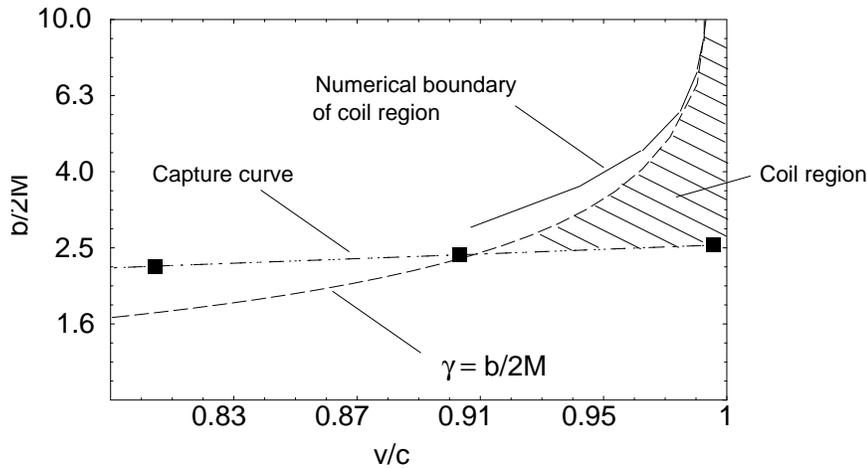}}
\caption{Coil formation region.}
\end{figure}

\subsection{String self-intersection}

The formation of coils is accompanied  by string self-intersection.
Self-intersection occurs if there exists a pair  of points on the
string worldsheet, $(\tau_1,\sigma_1)$ and  $(\tau_2,\sigma_2)$, such
that 
\begin{equation}\label{4.10}  
{\cal{X}}^{\mu}(\tau_1,\sigma_1)={\cal{X}}^{\mu}(\tau_2,\sigma_2)\, .
\end{equation}  
Since this condition forms a system of 4 equations for 4 variables, in
the  general case the self-intersection points are isolated. For the 
highly symmetric initial-value problem used here, a point of
self-intersection is  located at $Z=0$. 
%
%
Really, equation (\ref{4.10}) for $X_+$ given by (\ref{4.3}) 
implies that $\tau_1=\tau_2$. Since $Y(\tau,\sigma)$ given by
(\ref{4.4}) is symmetric with respect to $\sigma$ and as a function of
$\sigma$ has minimum at $\sigma=0$, the equation
$Y(\tau,\sigma_1)=Y(\tau,\sigma_2)$  has solutions
$\sigma_2=\pm\sigma_1$. Thus for a non-trivial solution of (\ref{4.10})
one must have 
\begin{equation}\label{4.10a}
(\tau_2,\sigma_2)=(\tau_1,-\sigma_1)\, .
\end{equation}
For definiteness, we assume that $\sigma_1> 0$. Since $Z(\tau,\sigma)$
given by (\ref{4.3a}) is an antisymmetric function of $\sigma$, the function
$Z$ vanishes at points of the string self-intersection.

To find a condition when the self-intersection of the string might occur
let us consider a function 
\begin{equation}\label{4.11} 
F(\zeta)=\zeta -2\gamma M\ln\left[1+(\zeta^2/Y_0^2)\right]\, .
\end{equation}
The point $Z=0$ is a point of self-intersection if 
the equation $F(\tau_1-\sigma_1)=F(\tau_1+\sigma_1)$ has a non-trivial
solution with $\sigma_1> 0$. This occurs
only when the function $F$ is non-monotonic, and the equation 
%
%
\begin{equation}\label{4.12} 
F'(\zeta) = 1-4\gamma M\zeta/(Y_0^2+\zeta^2)=0
\end{equation}
has a solution. The latter condition implies  $2\gamma M\ge Y_0$. This
is exactly the same condition of coil formation which was  obtained
earlier, Eqn.~(\ref{4.8}).

Suppose that $2\gamma M> Y_0$ and denote $\zeta_{\pm}=2\gamma
M\pm\sqrt{4{(\gamma M)}^{2}-Y_0^2}$. Then the function $F(\zeta)$ has a
local minimum at $\zeta=\zeta_+$ and local maximum at $\zeta=\zeta_-$.
For any value of $F_0$ between $F(\zeta_+)$ and $F(\zeta_-)$ there
exist 3 solutions of the equation $F(\zeta)=F_0$ which we denote by
$\zeta_1<\zeta_2<\zeta_3$. Correspondingly, there exist three different
branches of the solutions of the self-intersection conditions
\begin{eqnarray*}\label{4.12a}
\tau_1-\sigma_1&=&\zeta_1,\,\hspace{0.3cm}\tau_1+\sigma_1=\zeta_2\, ,\\
\tau_1-\sigma_1&=&\zeta_1,\,\hspace{0.3cm}\tau_1+\sigma_1=\zeta_3\, ,\\
\tau_1-\sigma_1&=&\zeta_2,\,\hspace{0.3cm}\tau_1+\sigma_1=\zeta_3\, .
\end{eqnarray*}
It is easy to verify that solutions from different branches have
different value of $\tau_1$. Transition between different branches take
place when either $\zeta_1=\zeta_2=\zeta_-$ or
$\zeta_2=\zeta_3=\zeta_+$.
Simple analysis shows that
the self-intersection is stable with respect to small perturbations of
the initial configuration of the string (before its scattering by the black
hole).

%
Coils are created and disappear at {\em cusps}. To shows this
let us study  the properties of vectors $X^{\mu}_{,\tau}$ and
$X^{\mu}_{,\sigma}$ which are tangent to the worldsheet of the string.
Note that $(X^{\mu}_{,\sigma})^2\equiv Y_{,\sigma}^2+Z_{,\sigma}^2\ge
0$ and hence vector $X^{\mu}_{,\sigma}$ is spacelike except for points
where it vanishes. It is easy to see that this occurs either at the
point of coil formation, $(\sigma=0, \tau=\tau_{-})$, or at the point
of its disappearance, $(\sigma=0, \tau=\tau_{+})$, where 
\begin{equation} 
\tau_{\pm}=2\gamma M \pm \sqrt{4(\gamma M)^2-Y_0^2}\, . 
\end{equation}
Constraint equations (\ref{2.4}) imply 
\begin{equation}
g_{\mu\nu}X^{\mu}_{,\sigma}X^{\mu}_{,\tau}=0\, ,   \hspace{0.5cm}
(X^{\mu}_{,\sigma})^2=-(X^{\mu}_{,\tau})^2\, . 
\end{equation} 
The second relation shows that $X^{\mu}_{,\tau}$ is everywhere timelike
except for two points where $X^{\mu}_{,\sigma}$ vanishes. It is easy
to verify that at these points $X^{\mu}_{,\tau}$ is null. Hence,
formation of a coil and its disappearance always occur at 
cusps. 

\subsection{Numerical results}

Figure~6 shows sections of the numerical worldsheet of  a string
propagating with an initial velocity of $0.995\,c$ ($\gamma = 10$) and 
impact parameter of $4\,{r}_{g}$. The view looking down onto the XZ
plane show a  complicated early phase associated with the evolution of
short-lived coils  and, at late times, two kink-like pulses 
propagating outward. The speed of  the pulses is again that of light.
The projections onto the YZ plane shows more clearly the early
evolution of the coils, which indicate that the points on the string
near the $Z = 0$ plane undergo a short-lived deflection  across the $Z
= 0$ plane. The formation of coils suggests that the behaviour  of the
string at periastron is particle-like in that there is insufficient
time  for tension to play a role, and the points on the string undergo
a temporary Keplerian deflection. However,  once clear of the black
hole, the tension can again assert itself and  solitonic pulses with an
S-shaped profile emerge and propagate outward.  Figure~6 also shows that the
string self-intersects in the $Z=0$ plane. As  the coil grows, the
intersection point moves gradually downward in  the Y-direction. Once
the coil has reached its maximum size, the  S-shaped kinks have formed
and begin to propagate outward, away from  the $Z = 0$ plane. The
intersection point disappears when  the kinks have completely
separated. 

\begin{figure}
\centerline{\epsfxsize=10cm \epsfbox{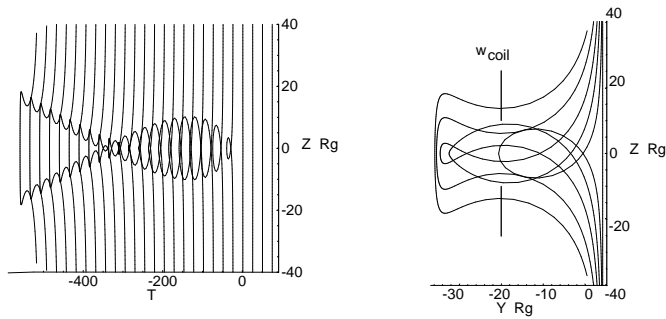}}
\caption{Time sequence of string scattering in ultra-relativistic regime
(numerical results). Black hole lies at origin of coordinate system. Initial 
velocity 0.995c, impact parameter 4.0 ${r}_{g}$.}
\end{figure}

Figure 7 shows YZ sections generated from the analytic weak-field and 
ultra-relativistic solutions for the same physical parameters and
intervals of proper  time as in Figure ~6. The two analytic solutions are 
virtually indistinguishable, the only difference being in the first
slice,  taken at the time where the string reaches periastron. The
ultra-relativistic solution  shows a completely straight string, as
expected, since the black hole does  not begin to distort the string
until it has passed over it. The weak-field  solution shows a slightly
bent string, indicating that even at $0.995 c$, the  influence of the
black hole is felt before closest approach. For later times,  the two
solutions are indistinguishable.

\begin{figure}[ht]
\centerline{\epsfxsize=10cm \epsfbox{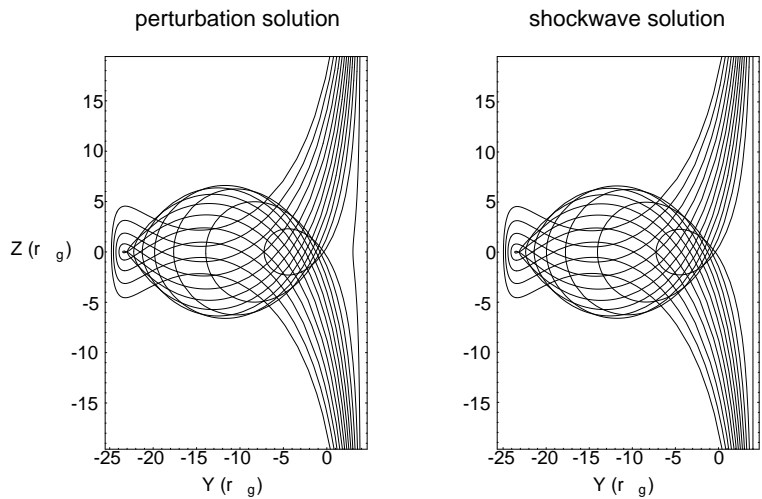}}
\caption{Time sequence of string scattering in ultra-relativistic
regime (analytic).  Black hole lies at origin of coordinate system.
Initial velocity, 0.995c, impact parameter 4.0 ${r}_{g}$.}
\end{figure}

As shown in Table~3, the maximum amplitude of the deflection and the size of 
the coils in the numerical data is slightly larger than predicted by 
Eqn.~(\ref{4.10}).
(The width of coils is difficult to estimate from numerical data, as is 
consequently shown with less precision that the other results.)
It is important to note that 
$\gamma = 10$ for these figures, which is low for the ultra-relativistic 
approximation, 
but represents a practical upper bound for numerical
computations\footnote{This is due to numerical anomalies that develop
when larger $\gamma$-factors are  considered. These numerical anomalies
manifest themselves as a rigid deflection of the entire string as it
crosses the $X=0$ plane. The cause of these anomalies was traced to a
loss of resolution in the first-order angular derivatives
${\partial}_{\phi}{\cal{X}^{\mu}}$ at large $\gamma$-factors. The
problem is suppressed by increasing the number  of grid points; the
large grid sizes required result in a numerically intensive solver.}.
Furthermore, the impact parameter is very small, and, as shown in the
previous section, the approximate solutions are no longer completely accurate
under these circumstances.

\begin{table}[ht]
\begin{center}
\caption{Scattering data - Schwarzschild ultra-relativistic.}
\begin{tabular}{lll}
\hline
 & & \\
 & numerical & perturbation \\
\hline
 & & \\
 $A_{\infty}$ &  $39.7\,{r}_{g}$ & $31.6\,{r}_{g}$ \\
 $w_{coil}$   &  $20\,{r}_{g}$ & $13.2\,{r}_{g}$   \\
 & &  \\
\hline
& ($v = 0.995\,c$, & $b = 4.00\,r_g$) \\
\end{tabular}
\end{center}
\end{table}

Since the weak-field and ultra-relativistic approximations are accurate
only  for impact parameters greater than $\sim 10 r_g$, the boundary of
the coil  formation is expected to shift due to strong field effects.
The solid line shown in Figure~5, based  on numerical tests to detect
coil formation, indicates that the boundary is  shifted towards lower
velocities; the strong field near the black hole tends  to enhance coil
formation\footnote{Our numerical results are inconsistent with the
findings of  Lonsdale and Moss\cite{LoMo:88}, who observe coils at much
lower velocities and larger impact parameters. 
%
%
The results discussed here clearly show that coil formation
is a high-velocity phenomenon. Other inconsistencies with \cite{LoMo:88} 
will be discussed in subsequent sections. }. 
%
%

The black hole drives the self-intersection process  as its
gravitational influence draws points on either side of the string 
towards the $Z = 0$ plane. 
%
%
What happens with the cosmic string forming a cusp which starts growing
into a closed coil cannot be answered in a simplified model we use. One
of the possibility is that after formation a small coil, as the result
of the string interconnection \cite{ShVi:94},\cite{AlYo:88}
the coil creates a loop. In this scenario,
the black hole driven self-intersection could leave a line of small
loops in the wake of the string. Another option is that loops never
form. Instead blobs of Higgs particles are ejected continuously as the
budding loop cuts from the string without fully forming. This would
result in a trail of fastly decaying Higgs particles behind this
cusp.\footnote{We are grateful to the referee who attracted our
attention to this possibility.}
%

\section{Near-critical Scattering}
\setcounter{equation}0

As we already mentioned, there are two possible outcomes when a cosmic string
interacts with a black hole: either the string is scattered, or it is captured 
by the black hole. For a
given initial velocity $v$ the outcome depends on the value of the
impact parameter $b$. In other words, in the space of initial data
$(v,b)$ there exists a critical line $b=b_{crit}(v)$ which separates
these two different regimes. We discuss the critical impact parameter
for string capture in Section~7. This section discusses string  
scattering where the impact parameter is
extremely close to the critical impact parameter for capture and the string 
comes close to the black hole. In a similar
situation for test particle scattering,  the particle can execute multiple 
orbits around the black hole before escaping. Some 
similarities with particle scattering are observed, but there are also 
important differences that shed light on the process of string
capture. 

Figure~8 compares a slice of the string worldsheet through the $Z=0$ 
plane to the motion of a test particle.  A comparison is made of a
series of string and particle trajectories with identical velocities
($v = 0.987 c$) and nearly identical impact parameters. For the
particle trajectories, these range from  $b = 2.60\,{r}_{g}$
through $b = 2.65\,{r}_{g}$. For the string, the impact
parameters are slightly smaller, ranging from  $b =
2.55\,{r}_{g}$ through $b = 2.56\,{r}_{g}$. The impact parameters
at the lower end of each range result in capture, whereas those
at the upper end result in scattering.
It is easy to see that,
even at ultra-relativistic velocities, tension influences string
motion. Although the critical impact parameter for the string
is very close to that for a particle, suggesting that tension
plays a limited role, the dynamics of the string are, nevertheless,
still governed by tension. 
%
%
This is manifested in the partial winding of the string while gravity
dominates the dynamics, followed by unwinding
of the string as a sufficient interval has elapsed for tension
to again dominate; the transition between the two regimes is
marked by the cusp-like feature in box B. By comparison, the escaping 
trajectories of the test particle are 
far less complicated, showing no folds or cusps.
%
%
%

\begin{figure}
\centerline{\epsfxsize=15cm 
\epsfbox{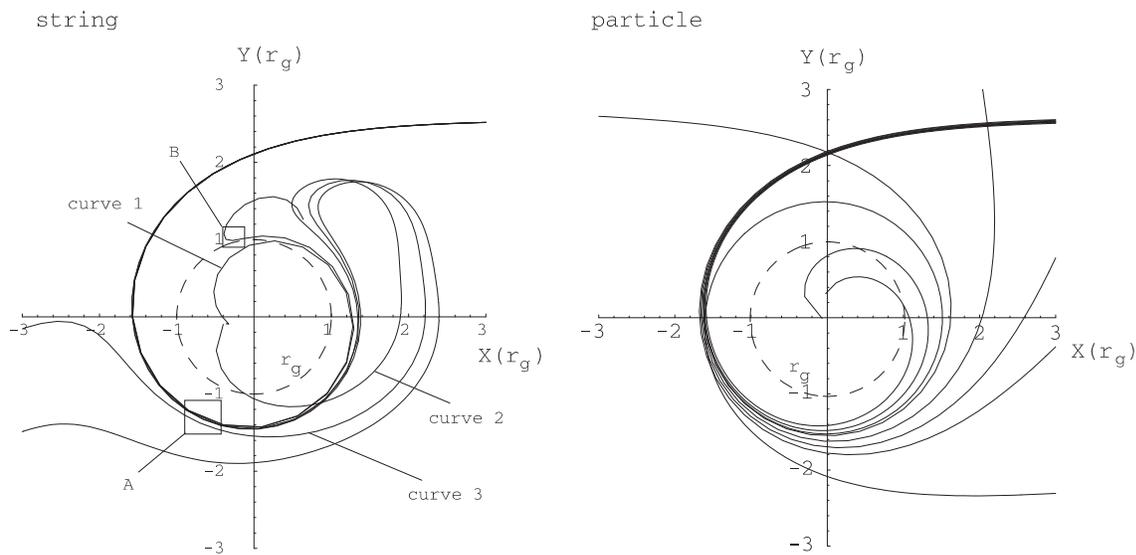}}
\caption{Near-critical scattering for string and test particle; $v =
0.987c$ and ${r}_{g} = 1$.}
\end{figure}

The detailed motion of the string in the critical regime close to
capture is highly
sensitive to the initial impact parameter. Figure~8 shows that there
are two types of capture trajectories, trajectories such as curve 1,
where the string crosses the horizon directly, and curve 2, where the
string folds back on itself before crossing the horizon. Curve 2
exhibits a coil-like feature, but does not represent a
self-intersection since the intersecting points have distinct proper
times. Escaping trajectories, such as curve 3, have a folded structure.
There are two  critical cases that mark the transition between each of
these three generic curves. The transition between curves of type 1 and
type 2 is marked by a structure  that develops an increasingly
cusp-like shape (point B in Figure ~8)  as the impact parameter approaches
the critical value that marks the transition from type 1 to type 2
curves. In practical terms, this impact parameter is difficult to
obtain since it requires a large number of significant digits (curve 1
and the curve with point B have impact parameters that differ by one
part in $10^5$). The transition between type  2 and type 3 curves is
marked by a tangent point, where the string trajectory passes twice
through the same point (point A) at different times  (again, no
self-intersection). Furthermore, the transition between type  2 and
type 3 curves is associated with the critical impact parameter for
capture.

A full $3D$ rendering of the worldsheet nearest the black hole can be 
generated, but it is very difficult to interpret and
is therefore not shown here. It is sufficient to note that
complicated folds and twists are generated as points on the string cross and
recross while executing their partial orbit of the black hole. At late times, 
however, these folds and twists have dissipated 
and all that remains are kinks, as shown in Figure ~9. 
%
%
This figure shows a detailed view near the $Z = 0$ plane; the kinks
evolve and move outward along the string, exiting at the top and bottom 
of Figure~9 and leaving behind the straight portion
visible at $Y \approx -15 {r}_{g}$.
%
%
This Figure
reinforces the idea  that scattering at late times is completely
understood in terms of the perturbative solutions.  However, it may
take a significant amount of time for the simple kink/anti-kink 
picture to emerge. In the case studied here, the distortions due to
the  close encounter persist until the string is about $1000\,r_g$ past
the black  hole. 

\begin{figure}[ht]
\centerline{\epsfxsize=7cm \epsfbox{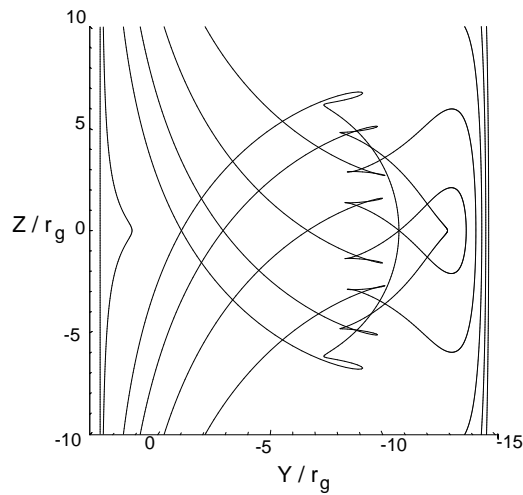}}
\caption{String worldsheet at late proper time, $v = 0.987 c$. Coils dissipate
and kinks emerge.}
\end{figure}

The numerical results for moderate $\gamma$-factors  show that the
string worldsheet folds back on itself, with no evidence of multiple
windings or glory scattering as would be the case for near-critical
scattering of particles.  In fact, it seems as if the transition to  a
"direct capture" trajectory (like curve 1 in Figure ~8) always occurs 
before a full turn is achieved.  The fact that multiple windings  are
absent is a clear indication that the tension in remote  parts of the
string eventually asserts itself in all cases accessible to the
numerical solver.  The question as to whether the string can complete more 
than one full turn and "wrap" the black hole for larger $\gamma$-factors
is an open one.

\section{Role of String Tension - Dust Strings}
\setcounter{equation}0

Since a cosmic string is an extended object under tension, motion of a 
string near a black hole represents the result of the competing
influences  of tension and gravity. In order to shed light on the role
of tension in the  dynamics of the string, the motion of the string is
compared to the motion  of an array of test particles initially 
configured with the same position and initial velocity as the cosmic
string.  To do this, consider a family of N test  particles arrayed on
a line with initial position ${{\cal{X}}^{\mu}}_{i}({\tau}_{0})$  and
initial velocity ${\partial}_{\tau}{{\cal{X}}^{\mu}}_{i}({\tau}_{0})$,
where $i$ is a position index (an integer between 1 and N) that
describes the  initial location of the test particle on the line. These
particles  each satisfy the geodesic equation and constraint,
\begin{eqnarray}\label{dust1}
{{d}^{2}\,{{\cal{X}}^{\nu}}_{i} \over d\,{\tau}^{2}}+
{\Gamma}^{\nu}_{\rho \sigma}\,
{d\,{{\cal{X}}^{\rho}}_{i} \over d\,\tau}\,
{d\,{{\cal{X}}^{\sigma}}_{i} \over d\,\tau}& = &0\, ,\\
\nonumber 
{d \over d \tau}\left[{g}_{\mu \nu}\,
{d\,{{\cal{X}}^{\mu}}_{i} \over d\,\tau}\,
{d\,{{\cal{X}}^{\nu}}_{i} \over d\,\tau}
\right] & = &0\, .
\end{eqnarray}
It is easy to show that, in flat spacetime,
\begin{equation}\label{dust2}
{X}^{\mu}_{i}(\tau) = \left(\cosh{(\beta)}\,\tau,
\sinh{(\beta)}\,\tau+{X}_{0},{Y}_{0},{\sigma}_{i}\right) 
\end{equation}
satisfies these equations and represents a straight dust string moving
with velocity $v = \tanh{\beta}$ where the discrete ${\sigma}_{i}$
mimic the $\sigma$ coordinate of the cosmic  string.

%
%
It should be emphasized that in the general case dust string dynamics is
quite different from the dynamics of the Nambu-Goto string. For example,
without tension a small bump on a dust string simply goes away, while a
similar bump on the Nambu-Goto string produces two small bumps which
propagate away from one another with the speed of light. Nevertheless,
we found it instructive to compare the motion of a straight Nambu-Goto string
in the field of a black hole with the motion of a similar dust string
since it clearly demonstrates the role of string tension in capture
and scattering processes. 
%
%

In the weak-field and ultra-relativistic cases, the role of tension in
the string is made apparent by comparing the YZ projections of Figures~2
and 6  against those of the dust string in Figure ~10. Figure 10
shows that  a coil always evolves in the dust string worldsheet. This
coil is due to the  Keplerian  nature of the trajectories of each
particle on the dust string. Particles  lying on one side of the $Z =0$
plane are deflected across this plane since  their motion is
constrained to an orbital plane passing through the black  hole. Unlike
the cosmic string, the Y-axis deflection of the dust string  is
unbounded (this is not to say that the deflection does not reach an 
asymptotic angular value). Close examination of Figures~2, 6, and 10
reveals that, at the early stages of scattering, the size
of  the coils in the dust and cosmic strings are virtually identical, 
reinforcing the idea that tension takes some time to assert itself.
Coil  formation is a generic feature of the dust string; this is not
the case  for the cosmic string, where coil formation is subject to
the  low-velocity cut-off  effect discussed above.

\begin{figure}
\centerline{\epsfxsize=10cm 
\epsfbox{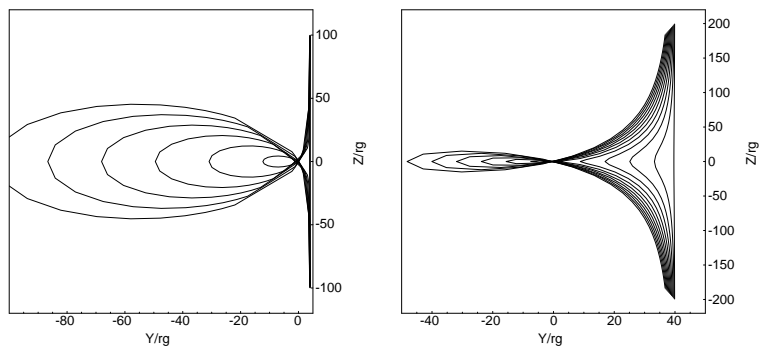}}
\caption{Time sequence of dust string scattering in ultra-relativistic
regime.  Black hole lies at origin of coordinate system. Left: Initial
velocity  0.995c, impact parameter 4.0 ${r}_{g}$. Right: Initial
velocity 0.76c,  impact parameter 40 ${r}_{g}$.}
\end{figure}

In the strong-field regime (with $v < c$), a comparison to the dust
solution is  uninformative  since the impact parameter is well below
the critical  value for particles and the portion of the dust string
near the equatorial  plane is captured by the black hole. However, it
does reinforce the idea  that internal tension plays an important role
in the dynamics of the string,  helping the string avoid capture for
impact parameters well below that of  the dust string. 

\section{String Capture}
\setcounter{equation}0

In Reference  \cite{DVFr:97} we presented a plot of the critical impact
parameter for capture as a function of velocity, the capture curve for
cosmic strings. In that paper, we used the massive-particle boundary
conditions exclusively and showed that, for string lengths greater than
$2000 r_g$, the capture curve was definitive for velocities greater
than $0.2 c$.  We are now in a position to revisit this curve using the
perturbative boundary conditions in the improved numerical solver, and 
to merge the numerical findings with the ultra-low velocity results of
\cite{DVFr:98} to present a complete capture curve.

\begin{figure}[ht]
\centerline{\epsfxsize=10cm 
\epsfbox{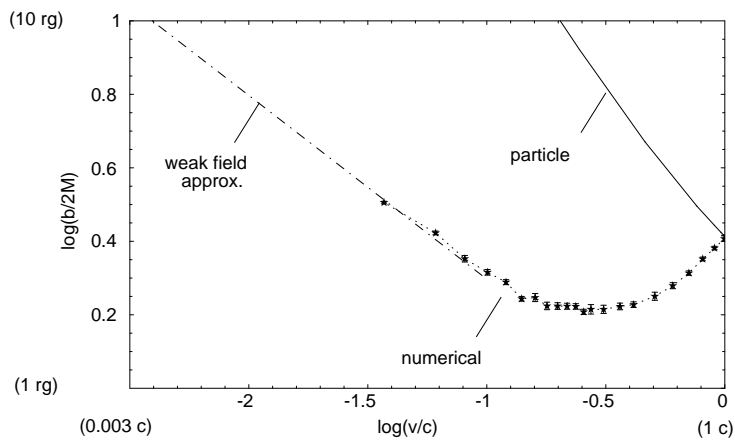}}
\caption{Cosmic string capture curve for Schwarzschild black hole - data from
numerical and perturbative results.}
\end{figure}

The revised capture curve shown in Figure~11 validates two claims made 
in the previous paper. First, that the capture curve for long strings 
is indeed definitive (provided focusing of end points is negligible).
This  is so because the revised capture curve is identical to  that of
the $L = 2000\,r_g$ string for $v > 0.2 c$.  Second, that there is
indeed a minimum impact parameter at  intermediate velocities, which
occurs at $v \approx 0.2 c$  and has a value of $1.6\,
r_g$.  However, it is important to note that the new solver is also
velocity-limited in that, at sufficiently low velocities, second-order
perturbative effects become important and invalidate the first-order
solutions used to set the boundary conditions. Nevertheless, analytic
results \cite{DVFr:98,Page:98}  are available for  this velocity range, and the
numerical results join smoothly with the analytical result for the
critical impact parameter for very low velocities
\begin{equation}\label{n6.1}  
b_{capture} \sim {r_g \over 2} \sqrt{{\pi \over 2 v}}\, ,
\end{equation}
and is shown as a dashed line in Figure ~11. This capture curve also is in
a good agreement with a recent result by Page \cite{Page:98} for the
critical impact parameter,
\begin{equation}\label{n6.1a}  
b_{capture} \sim {r_g \over 2} \left[\sqrt{{\pi \over 2 v}}-
\left(\sqrt{{\pi \over 2}}+{64 \over 15}-\sqrt{27} \right) +
{64 \over 15} v \right]\,.
\end{equation}
This formula reproduces the general shape of the curve shown in Figure  11,
without the fine structure that is revealed by the numerical approach. 
The error bars \footnote{As defined in \cite{DVFr:97}, the
lower error bar denotes the largest impact parameter resulting in capture
and the upper error bar denotes the smallest impact parameter resulting in  
escape for a given initial velocity.} in Figure ~11 are quite small. 
%
%
This suggests that the capture curve is not a smooth line, but
has some fine structure to it; this is particularly obvious at
lower velocities, but this structure is also observed on a smaller
scale at higher velocities as well. This could possibly be explained by 
the complicated
nature of the trajectory of the string for near-critical impact 
parameters, and its extreme sensitivity to initial data. However,  
detailed studies
(carried out subsequent to 
the drafting of the present paper)
of the capture
curve at ultra-relativistic velocities reveal that the structure
is damped out to a certain extent by using finer grids, so it is important
to note that some (and perhaps all) of the "jitter" in Figure~11 may be 
numerical in origin. 
%
%

It should be emphasized that our numerical results demonstrate that in
the limit $v\rightarrow c$ the critical impact parameter reaches the
same value, $3\sqrt{3}GM/c^2$, as the critical impact parameter for
capture of ultrarelativistic particles. This result is inconsistent
with the earlier results of Moss and Lonsdale \cite{LoMo:88}. Moreover,
the analytical result (\ref{n6.1}) also  is inconsistent with the 
results of Moss and Lonsdale, who quote a low-velocity dependence of
${v}^{-1}$ obtained from a numerical fit to their
capture curve.

\section{Conclusions}
\setcounter{equation}0

In this paper, we discussed the gravitational scattering of cosmic
strings by Schwarzschild black holes. This paper brought together
earlier results for the critical impact parameter for capture,
weak-field analytic solutions, and studied the limitations of
approximate solutions as a function of velocity and impact  parameter
by generating numerical solutions to the equations of motion. Earlier
findings have been corroborated by these latest  results, and 
numerical, weak-field, and shockwave results are consistent with one
another in the regions where they overlap. 

The scattering problem was studied in three regimes.  In the
ultra-relativistic regime, we showed that strings form coils, provided
that the conditions $Y_0 \le 2\gamma M$ and $Y_0 > b_{capture}$  are
satisfied; this phenomenon cuts off for velocities below $\sim 0.9 c$.
A detailed comparison of analytic and numerical results showed that the
perturbative results break down for $Y_0 < 10 r_g$; this is a
surprisingly small value and suggests that the weak-field approximation
is quite acceptable for all cases except near-critical scattering.
Where the weak-field approximation breaks down, in the strong-field
regime,  kinks are produced (without coils for $v < 0.9 c$, with coils
for $v > 0.9 c$), but their amplitude is larger than predicted by 
approximate solutions.

In the case of near-critical scattering, the string worldsheet evolves
highly complicated structures. A detailed  investigation of string
motion in this case showed that there are two characteristic
trajectories for capture, the first where the string enters directly,
the second where the string evolves a small coil before crossing the
horizon. This complicated behaviour, and its extreme sensitivity to
%
%
initial data, may help explain the structure observed in the capture curve;
however, a numerical origin to this structure has not been ruled out.
%
%

This paper has also shown the critical role
performed by tension in determining the dynamics of the string. The
absence of multiple  string windings around the black hole and the
complicated folds observed in the string worldsheet at near-critical
scattering, even at moderate $\gamma$-factors, 
%
%
are consequences of internal tension.
%
%
%

\vspace{12pt} 
{\bf Acknowledgments}:\ \ This work was partly supported by the Natural
Sciences and Engineering Research Council of Canada.  One of the
authors (V.F.) is grateful to the Killam Trust for its  financial
support.  The authors are also grateful to Don Page for numerous
discussions.

\appendix
\section*{Appendix - Analytical Results in Shockwave Regime}
\renewcommand\theequation{A.\arabic{equation}}
\setcounter{equation}0

In this Appendix we present a solution for an ultra-relativistic
straight string scattered by a black hole. The key observation which
leads to a solution of the equations of motion in the
ultra-relativistic limit is the following: in the reference frame of
the string the black hole moves with $v\approx 1$ and its gravitational
field is boosted to the shock wave \cite{HaSa:94}.  As a result, before
and after crossing the null surface $N$ representing  the black hole,
the string obeys the free equations in flat spacetime. All the
information concerning the non-linear interaction with the
gravitational field of the black hole can be obtained in the form of
``jump'' conditions on the null surface $N$. Such solutions were
studied earlier (see e.g. Reference ~\cite{AmKl:88}). 

To obtain the metric of an ultra-relativistic black hole one starts with
the metric (\ref{2.9}) written in the form
\begin{equation}\label{n8.1} 
ds^2=(1+2\Psi)ds_0^2+2(\Phi+\Psi)dT^2\, ,
\end{equation}
where
\begin{equation}\label{n8.2} 
ds_0^2=-dT^2+dX^2+dY^2+dZ^2\, ,
\end{equation}
and make the boost transformation
\begin{equation}\label{n8.3} 
\bar{T}=\gamma(T-vX)\, ,\hspace{0.5cm}
\bar{X}=\gamma(X-vT)\, ,\hspace{0.5cm}
\bar{Y}=Y\, ,\hspace{0.5cm}
\bar{Z}=Z\,  .
\end{equation}
Let $X_{\pm}=\bar{T}\pm\bar{X}$ so that
\begin{equation}\label{n8.3a}
X_{+}=\gamma(1-v)\,(T+X)\, ,\hspace{0.5cm}
X_{-}=\gamma(1+v)\,(T-X)\, , 
\end{equation}
and, since $\gamma^2 = {\left( 1 - v^2\right)}^{-1}$,
\begin{eqnarray}\label{n8.3b}
T & = & {\gamma \over 2} (1+v)\,
\left(X_{+} +\left({1 - v \over 1 + v}\right) X_{-}\right)\, ,\\
\nonumber
X & = & {\gamma \over 2} (1+v)\,
\left(X_{+} -\left({1 - v \over 1 + v}\right) X_{-}\right)
\, . 
\end{eqnarray}
Using Eqns.~(\ref{n8.3b}) the metric (\ref{n8.1}) takes the
form 
\begin{eqnarray}\label{n8.4} 
ds^2 & = & (1+2\Psi)\left[-dX_-\, dX_+ +dY^2+dZ^2\right] \\
\nonumber & + & {\gamma^2 \over 2}(1+v)^2
(\Phi+\Psi){\left(dX_+ +{1-v \over 1+v} dX_-\right)}^2\, .
\end{eqnarray}

In the lowest order $\Phi=\Psi=\varphi = M/R$, where $R=\sqrt{X^2+Y^2+Z^2}$ 
can be rewritten using Eqn.~(\ref{n8.3b}) as
\begin{equation}\label{n8.5a} 
R = \gamma\,\sqrt{{\left({1+v \over 2}\right)}^2\,
{\left(X_{+} -\left({1 - v \over 1 + v}\right) X_{+}\right)}^2+
{1 \over \gamma^2} \left(Y^2+Z^2\right)}
\end{equation}
so that
\begin{equation}\label{n8.5} 
\varphi={\gamma M
(1-v^2)\over\left[\left( {1+v\over 2}\right)^2
\left(X_+ -\left({1-v\over 1+v}\right)X_-\right)^2+
(1-v^2)(Y^2+Z^2)\right]^{1/2}}
\, .
\end{equation}

The metric (\ref{n8.4}) takes the form 
\begin{eqnarray}\label{n8.4a} 
ds^2 & = & -dX_-\, dX_+ +dY^2+dZ^2 \\
\nonumber & + & 2\varphi\left(dY^2+dZ^2\right) + 
\gamma^2 \varphi\left((1+v)^2\,d X_+^2 + (1-v)^2 dX_-^2\right)
\end{eqnarray}

In the limit $v\rightarrow 1$ and $\gamma M$ fixed one has \cite{AmKl:88}
\begin{equation}\label{n8.6} 
\lim_{v\rightarrow 1} \gamma^2\varphi=
\lim_{v\rightarrow 1}{\gamma M\over \sqrt{X_+^2+(1-v^2)\rho^2}}=-\gamma M
\delta(X_+)\ln \rho^2
\, ,
\end{equation}
where $\rho^2=Y^2+Z^2$. From this it follows that the metric in the
Aichelburg-Sexl form \cite{AiSe:71} is obtained
\begin{equation}\label{n8.7} 
ds^2=-dX_-\, dX_+ +dY^2+dZ^2 -4\gamma M F \delta(X_+) dX_+^2
\, .
\end{equation}
where $F=\ln \rho^2$.
The boosted metric in this form represents a gravitational shockwave.
The following components of the Christoffel symbols do not vanish:
\[
\Gamma^-_{++}=4\gamma M\, F \,\delta '(X_+)\, ,
\hspace{0.5cm}
\Gamma^-_{+Y}=2 \Gamma^Y_{++} =4\gamma M\, F_{,Y}\, \delta (X_+)\, ,
\]
\begin{equation}\label{n8.8} 
\Gamma^-_{+Z}= 2 \Gamma^Z_{++} =4\gamma M \, F_{,Z}\, \delta (X_+)\,  .
\end{equation}

The in-coming straight string motion (\ref{2.5}) 
in this limit takes the form 
\begin{equation}\label{n8.9} 
\left. X_+\right|_{{\tau}<0}={\tau}\, ,\hspace{0.5cm}
\left. X_-\right|_{{\tau}<0}={\tau}-2 \gamma X_0\, ,\hspace{0.5cm}
\left. Y\right|_{{\tau}<0}=Y_0\, ,\hspace{0.5cm}
\left. Z\right|_{{\tau}<0}=\sigma
\,.
\end{equation}

First solve the string equation (\ref{2.3}) in the metric
(\ref{n8.7}) for the initial conditions (\ref{n8.9}) using the
conformal gauge in which $\sqrt{-h}h^{AB}=\eta^{AB}$, so that
$\Box=-\partial^2_{{\tau}}+\partial^2_{\sigma}$. 

The equation for $X_+$ takes a simple form
\begin{equation}\label{n8.10} 
\Box X_+=0 \, .
\end{equation}
Hence its solution obeying the proper initial conditions is
\begin{equation}\label{n8.11} 
X_+={\tau} \, .
\end{equation}

The equation for $Y$ is
\begin{equation}\label{n8.12} 
\Box Y=2\gamma M\, F_{,Y}\, \delta({\tau}) \, .
\end{equation}
The solution obeying the initial conditions (\ref{n8.9}) is
solved using Green's Function methods \cite{DVFr:98} and
has the form
\begin{equation}\label{n8.13} 
\left. Y\right|_{{\tau}>0}=Y_0-2\gamma M\, 
\left[\arctan\left({{\tau}+\sigma\over Y_0}\right)
+\arctan\left({{\tau}-\sigma\over Y_0}\right)
\right] \, .
\end{equation}
Similarly, solving equation
\begin{equation}\label{n8.14} 
\Box Z=2\gamma M\, F_{,Z}\, \delta({\tau}) \, 
\end{equation}
for $Z$ with the initial conditions (\ref{n8.9}) one gets
\begin{equation}\label{n8.15} 
\left. Z\right|_{{\tau}>0}=\sigma-\gamma M\,\ln 
\left[{Y_0^2+({\tau}+\sigma)^2\over Y_0^2+({\tau}-\sigma)^2}
\right] \, .
\end{equation}

Expression for $X_-$ can be easily obtained by using
 constraint
equations (\ref{2.4}) which for the case under consideration take the
form ($\tau>0$)
\begin{equation}\label{n8.18} 
\dot{Y}^2+{Y'}^2+\dot{Z}^2+{Z'}^2=\dot{X}_- \, ,
\end{equation}
\begin{equation}\label{n8.19} 
2(\dot{Y}{Y'}+\dot{Z}{Z'})={X'}_- \, .
\end{equation}
Here a dot and prime denote derivatives with respect to ${\tau}$
and $\sigma$.
By integrating these equations one gets
\[
\left.X_-\right|_{{\tau}>0}=\tau -2\gamma X_0 -
2\gamma M \left[\ln \left( 1+\left(\tau+\sigma\over
Y_0\right)^2\right)+\ln \left( 1+\left(\tau-\sigma\over
Y_0\right)^2\right)\right]
\]
\begin{equation}
+
{8\gamma^2 M^2\over Y_0}\left[ \arctan\left( \tau+\sigma\over Y_0\right)+
\arctan\left( \tau-\sigma\over Y_0\right)\right]\, .
\end{equation}

\end{document}